\documentstyle[aps,pra,eqsecnum,preprint]{revtex}
\def\bea{\begin{eqnarray}}
\def\eea{\end{eqnarray}}
\def\be{\begin{equation}}
\def\ee{\end{equation}}
\begin{document}
\tightenlines
\author{Krzysztof Sacha and Jakub Zakrzewski}
\address{
 Instytut Fizyki imienia Mariana Smoluchowskiego,
  Uniwersytet Jagiello\'nski,\\
 ulica Reymonta 4, PL-30-059 Krak\'ow, Poland
}
\title{H atom in elliptically polarized microwaves:\\
Semiclassical versus quantum resonant dynamics
}
\date{\today}
\maketitle
\begin{abstract}
The dynamics of Rydberg states of atomic hydrogen
illuminated by resonant elliptically
polarized microwaves is investigated both semiclassically and
quantum mechanically in a simplified two-dimensional model of
an atom. Semiclassical predictions for  quasienergies of the
system are found to be in a very good agreement with exact quantum
data enabling a classification of possible types of
motion and their dynamics with the change of
 the ellipticity of the microwaves. Particular attention is
paid to the dynamics of the nonspreading wave packet states
which are found to exist for an arbitrary microwave polarization.

\end{abstract}
\narrowtext
\section{Introduction}

Consider a hydrogen atom, initially in a high Rydberg state with the
principal quantum number $n_0$ illuminated by the microwave field
of a frequency $\omega$ close to the frequency of the unperturbed Kepler
motion $\omega_K=1/n_0^3$. Quantum mechanically speaking, a resonant
periodic field couples strongly several $n$ states due to
 almost resonant transitions
$n\rightarrow n'=n\pm 1$ for $n$ close to $n_0$. Since the driving field
is periodic, applying the Floquet theorem \cite{s65} one may find eigenstates
of the atom-field system (the so called Floquet
or dressed \cite{cct} states). The eigenenergies are then referred to as
quasienergies of the system and are defined modulo $\hbar\omega$.
The Floquet states, periodic in time, may be viewed as linear
combinations of unperturbed system eigenstates. While the construction
of Floquet states is possible both for nonresonant and resonant
driving,  in the latter case they may have quite unusual properties,
especially in a semiclassical limit.

Classically a resonance between the driving frequency and the frequency of the
unperturbed motion leads to a strong perturbation of the system and a creation
of a stable island in the phase space centered on a periodic orbit
of the frequency $\omega$. The motion in the island is locked to the microwave
frequency due to the nonlinear resonance \cite{lil}. Semiclassically one then
expects that the corresponding Floquet time-periodic state will follow
a classical trajectory (in the vicinity of the periodic orbit) i.e.
form a wavepacket which will not disperse in time.

States localized in the resonance island for such a periodic perturbation
have been first considered more than 20 years ago \cite{beza77} and details
of their semiclassical construction for some one-dimensional (1D) model systems
have been analyzed \cite{holt,sk95}. The wavepacket character of the time 
evolution
of individual Floquet states has been realized only quite recently \cite{ab3}
for hydrogen atom driven by linearly polarized microwaves. Independently,
it has been shown that Gaussian wave packets may propagate almost
without dispersion along circular periodic orbits
in hydrogen atoms driven by circularly polarized fields \cite{ibb}.
The fact that the harmonic approximation implied in \cite{ibb} and resulting
the Gaussian wave packet form is not a necessary condition for non-spreading
properties have been discussed in \cite{dzb95,zdb95} where it was shown
that the harmonic expansion provides a good approximation for exact
Floquet states of the system (which by definition, being time periodic, do
not spread on a long time scale).

The fact that one may construct in a nonlinear system wavepackets that
do not spread induced a flurry of activity in the field. Some 
\cite{far_c,ibb_r,fluA95,bufL96,lbf,clfu97}
concentrated on modifying the potential so as to make the harmonic
approximation as good as possible aiming at the construction of Gaussian
nonspreading wavepackets. Claiming that the anharmonic corrections are
big for a hydrogen atom driven by circularly polarized fields those authors
added a magnetic field perpendicular to the microwave polarization plane
\cite{far_c,ibb_r,fluA95,bufL96,lbf,clfu97}. 
This helps to minimize the unharmonicities {\it in this plane}
leaving, however, unaffected the terms along the magnetic field axis.
Thus the Gaussian wave packets remain even then an approximation to the
real dynamics and must disperse (although very slowly in time). Another
approach aimed at optimizing the coordinate system to the symmetry of the
problem \cite{keb}.

As discussed by us elsewhere \cite{dzb95,zdb95,szd98,zdb98,jasz98,dz98} much more
fruitful is another approach, already outlined above. Namely we define
the nonspreading wavepacket as a single Floquet state (for which the
Gaussian packet may be merely an approximation). Then a localization of
the wavepacket in the vicinity of a stable fixed point is assured by the
correspondence principle provided the size of the surrounding island
in the phase space is comparable to $\hbar$. One may construct
an approximate resonance Hamiltonian in the vicinity of the island
whose eigenstates will approximate well those Floquet states which
are localized in the vicinity of the island (see below).
 Simultaneously, time-periodicity
of Floquet eigenstates assures that the exact Floquet states will not disperse.

The existence of such wave packet Floquet eigenstates has been proven by
an exact numerical diagonalization of the problem both for linear 
polarization (LP) \cite{ab3}
and circular polarization (CP) \cite{dzb95,zdb95} of the microwaves. To
allow their detection one should consider the ways of populating
effectively such states. For a CP, where the direct optical
excitation from a weakly perturbed low lying state is impossible (since
the wavepacket is built from predominantly circular atomic states
unaccessible from low lying states due to dipole selection rules) one
should first prepare the atomic circular state \cite{kl,dd} and then switch on
the microwaves sufficiently fast \cite{dzb95}. It has been shown that
the wavepacket states may be populated in this way with about 90\% efficiency
\cite{zd97}. For a LP case, an addition of a static electric
field allows to control the trajectory of the wavepacket \cite{szd98}, in
particular wavepackets moving along elongated, 
low angular momentum trajectories may be created.
 Such wavepackets may be accessible to a direct
excitation from low lying states.

The next important question is the possible mechanism of the detection of these
states. At least  three possible ways suggest themselves. Two of them
utilize the residual decay of wavepackets, 
either via the spontaneous emission (treated both
for LP \cite{hb98} and CP \cite{zibb97,dz98}) or the ionization. The
former may not be efficient, since, at least for wavepackets moving on circular
trajectories, the corresponding spontaneous emission rates are quite small
\cite{dz98}. In the ionization experiment, 
the population of the wavepacket state may be detected by a strong decrease in
the ionization yield (since ionization rates of wavepacket states
are typically very small \cite{dzb95,zdb95}). On the other hand, 
these rates  fluctuate strongly  (the mechanism of
their ionization, via ``chaos assisted tunneling'' is discussed in detail
elsewhere \cite{zdb98,jasz98}) -- this may make their detection 
in the ionization yield
quite ambiguous.

By far the most promising method is the Floquet spectroscopy \cite{ab4}
i.e. probing, by a second weak microwave field, the structure of Floquet
(dressed \cite{cct} by microwaves) states. To this end a precise estimation
of the quasienergies of wavepacket states is necessary. An exact
diagonalization of the problem gives all the Floquet states and a time
consuming inspection of individual eigenvectors is necessary to identify
the wave packet states.
This process may be optimized by calculating properties of
matrix elements of appropriately chosen operators but certainly it is desirable
to have good semiclassical predictions for the quasienergies. For CP case
those are given, to a very good accuracy from the harmonic approximation
Hamiltonian \cite{ibb,dzb95,zdb95}, this approach being, however, restricted
to this particular system.

For a general case of periodically driven system there is no simple unitary
transformation which removes the time-dependence (as it is in the CP case)
and the correct approach is to use approximate resonant Hamiltonians. The
semiclassical quantization of such a Hamiltonian gives not only  the
good estimate for wavepacket states but allows for the classification of
resonant states for systems of more than one dimensions. Recently using
such an approach we could discuss the resonant dynamics in a realistic 
three dimensional (3D)
model of a hydrogen atom in the LP case \cite{bsdz98}. Similarly, we
have discussed the control of wavepacket trajectories using an
additional static electric field \cite{szd98}.

Up till now the discussion of nonspreading wavepacket states has been
restricted to linear and circular polarization cases only. The aim of this paper
is to treat a resonant dynamics of a hydrogen atom, both semiclassically and
quantum-mechanically in a general case of an elliptical polarization (EP).
Apart from generalizing the notion of nonspreading wavepackets to an
arbitrary EP, we discuss the full dynamics
of quasienergies as a function of the ellipticity of the microwaves for
a resonant case, being
 stimulated by recent
ionization experiments \cite{koch96}. Unfortunately, the experiments do not
allow for a full selection of the initial state of an atom (states with
all possible angular momenta are simultaneously excited) but this situation
may improve in the near future.

The EP case is highly nontrivial.
For LP microwaves the conservation of
the angular momentum projection onto the polarization axis, $L_z$, makes
the dynamics effectively two-dimensional (2D). For the CP case
while $L_z$ is not conserved, the transformation to the frame rotating
with the microwave frequency removes the explicit oscillatory
time-dependence (see e.g. \cite{nauen2,How92,zdgr93,ZGD96}).
 Both these simplifications are no longer
possible in the general EP microwave field. In effect, the exact
quantum diagonalization approach for the part
of the spectrum corresponding to the strongly perturbed atomic Rydberg
spectrum would require very big computer memory.
  Let us mention also that in the effective
2D LP case, the state of the art computations \cite{ab3} consider
initial atomic states with the principal quantum number of the order of
20.

For that reason we shall consider not the realistic fully 3D model
of an atom but rather the restricted 2D model in which the electronic
motion is restricted to the polarization plane. Study of such
simplified models have been most successful in the past both for
LP (where one dimensional model has been a main source
 of quantum results for
a long time \cite{JS91,KoLe95,sk95}) and in CP \cite{nauen2,How92,zdgr93,ZGD96}
 where also the 2D, polarization plane restricted model has been
utilized. Additional argument favoring the 2D model comes from our
classical study of dynamics in EP microwaves \cite{sz97,sz98}
where comparison of 2D \cite{sz97} and 3D \cite{sz98} analysis
shows the similarity of physical phenomena in both cases. Simply
put the perturbation is most effective if the polarization plane
coincides with the plane of Kepler motion.

\section{The semiclassical versus quantum approaches}

The Hamiltonian of the hydrogen 2D model atom
driven by an elliptically polarized electromagnetic field reads
in the dipole approximation and
in the length gauge (in atomic units)
\be
        H=\frac{p_{x}^{2}+p_{y}^{2}}{2}
        -\frac{1}{r}+ F(
        x\cos{\omega t}+\alpha y\sin{\omega t}),
\label{ham}
\ee
where $r= \sqrt{x^2+y^2}$ while $F$ and $\omega$ denote
 the amplitude and the frequency of the microwave
field, respectively.
$\alpha$ defines the ellipticity of the microwaves with
$\alpha=0$ ($\alpha=1$) corresponding to a LP (CP) limiting case.

Using the Floquet theorem the solution of the quantum problem
is equivalent to diagonalizing the Floquet Hamiltonian
\be
\left ( H-i\frac{\partial}{\partial t}\right ) \psi_n =H_F\psi_n=
 E_n\psi_n
\label{floq}
\ee
with $E_n$ being the quasienergies while $\psi_n$ time-periodic
Floquet eigenstates.

The details of the numerical method are described in the Appendix.
In short, the calculations proceed by expressing the
Floquet eigenvalue
equation in the scaled semi-parabolic variables
\bea
x=\Lambda\frac{u^2-v^2}{2}, & y=\Lambda uv.
\label{scaluv}
\eea
where $\Lambda$ is an arbitrary scaling factor.
This allows to remove the Coulomb singularity and cast the Schr\"odinger
equation into the generalized eigenvalue problem for
polynomial like operators. Standard harmonic oscillator
creation and annihilation operators
allow for simple evaluation of matrix elements.
 The approach closely
resembles that for the 2D atom in the CP field \cite{ZGD96} except that
the explicit time dependence is treated by the Fourier expansion. It
is worth stressing that using the complex scaling parameter $\Lambda$
one effectively realizes the complex rotation in the system which
enables the exact treatment of the coupling to the continuum (ionization),
for details see \cite{ab3,ZGD96} for LP and CP cases, respectively.
In this paper, however, we shall concentrate on the Floquet level dynamics only
for the sake of the comparison with the semiclassics. 
The analysis of the ionization
phenomenon is left for a future publication.

The semiclassical quantization of resonant dynamics closely resembles
the similar procedure applied by us recently for the LP case \cite{szd98,bsdz98}.
That in turn originates from a general prescription for EBK quantization
of Floquet spectra \cite{BrHo91}.

Starting with the Hamiltonian (\ref{ham}) we remove the explicit time dependence
by going to the extended phase space \cite{lil}.
 Defining the momentum $p_t$
conjugate to $t$ (time) variable we get the new Hamiltonian
\be
{\cal H}=H+p_t
\label{ext}
\ee
which is conserved during the motion.
The quasi-energies of the system will be then
 the quantized values of ${\cal H}$.

As the next step we express the Hamiltonian
 in action-angle variables of the unperturbed Coulomb problem \cite{sz97}.
 For the 2D model atom those are e.g.
 the canonically conjugate
pairs $(J,\theta)$ and $(L,\phi)$. $J$
is the principal action
 (corresponding to the principal quantum number, $n_0$).
The corresponding angle, $\theta$, determines the position of the
electron on its elliptic trajectory.
$L$ is the angular
momentum (equal to $L_z$ for the 2D motion in the $x-y$ plane)
 while $\phi$
is the conjugate angle (the angle between the Runge-Lenz vector
and the $x$-axis, i.e. the main axis of the polarization ellipse).

We shall consider below the case of the resonant driving,
 i.e. when the frequency
of the Kepler motion $\omega_K=1/J^3$ is close to the microwave driving frequency
$\omega$. Applying the secular perturbation theory \cite{lil}
to average over the nonresonant
terms one obtains an approximate resonant Hamiltonian (in the frame
rotating together with the electron) of the form
\be
{\cal H}_r=-\frac{1}{2J^2}-\omega J+
F\Gamma(L,\phi;\alpha)\cos(\hat\theta-\delta)+\hat p_t
\label{EPr}
\ee
where $\hat\theta=\theta-\omega t$ while $\hat p_t=p_t+\omega J$.
${\cal H}_r$ yields the pendulum like principal action motion
with the strength and the equilibrium position determined by 
$\Gamma(L,\phi;\alpha)$ and $\delta=\delta(L,\phi;\alpha)$ respectively.
The pendulum Hamiltonian is obtained by additionally expanding ${\cal H}_r$
around the center of the resonance island given by $J=\omega^{-1/3}$
up to the second order in $\Delta J=J-\omega^{-1/3}$ but we do not apply this
expansion.
Both $\Gamma$ and $\delta$  
depend on the initial shape and orientation of the electronic ellipse
(via $L,\phi$) as well as on the ellipticity of microwaves, $\alpha$,
and are given by \cite{sz97,sth}
\begin{equation}
\Gamma(L,\phi;\alpha)
=\left[\left(\frac{1+\alpha}{2}V_{1}
\right)^{2}+2\cos 2\phi \frac{1+\alpha}{2}V_{1}\frac{1-\alpha}{2}V_{-1}
+\left(\frac{1-\alpha}{2}V_{-1}\right)^{2}\right]^{1/2},
\label{gam}
\end{equation} 
\begin{equation}
\tan\delta=\frac{(1-\alpha)V_{-1}-(1+\alpha)V_1}{(1-\alpha)V_{-1}+(1+\alpha)V_1}
\tan\phi,
\end{equation}
where
\begin{equation}
 V_{\pm 1}(J,L) =\omega^{-2/3}[{\cal J}_{1}^{'}(e) \pm
 \frac{\sqrt{1-e^2}}{e}{\cal J}_{1}(e)].
\end{equation}
${\cal J}_{1}(x)$ and ${\cal J}_{1}^{'}(x)$ denote the ordinary 
Bessel function and its derivative, respectively, while 
$e=\sqrt{1-L^2/J^2}=\sqrt{1-L^2\omega^{2/3}}$ stands for an eccentricity of 
an electronic ellipse.
For completness let
us mention that a similar approximation on the purely quantum level,
leading to Mathieu equation
has been performed for the CP case only in \cite{ke96}.

The semiclasical quantization of (\ref{EPr}) is straightforward and
follows closely the procedure described in detail elsewhere
\cite{bsdz98} for arbitrary $m:1$ resonance. For $1:1$ resonance
considered here, the trivial
quantization of
 $\hat p_t$,  exploring the time periodicity of the system,
 yields additive terms
$k\omega$ to quasi-energies (different values of $k$
correspond to  different Floquet zones).
As discussed first in \cite{lr86}, see also \cite{bsdz98}, the
orbital motion
 in $(J,\hat \theta)$ variables
 (along the perturbed Kepler ellipse)
  is much     faster than the modification of the ellipse shape and
its movement (precession) as described by the motion in $(L,\phi)$.
Thus the slow and fast motions may be adiabatically decoupled.
Making Born-Oppenheimer approximation and using standard WKB rules,
the fast $(J,\hat\theta)$ motion is quantized taking the
Maslov index $\nu=2$ (corresponding to librations, i.e.,
 we quantize states inside the resonance island).
  Being interested in resonantly
localized states we shall consider later the
ground
state of the radial motion, only.
For the slow angular $(L,\phi)$ motion
we take the Maslov index  $\mu=0$ or 2 for a rotational
or librational motion, respectively.

Similarly, as in the LP microwaves \cite{bsdz98},
 it is easier to quantize first the
slow motion
generated by constant values of $\Gamma(L,\phi;\alpha)$ and later
treat the fast motion.  Such a procedure is justified since 
quantizing the fast motion one takes $\Gamma(L,\phi;\alpha)$ as a constant
quantity and thus  the order of the quantization
does not matter.

The existence of the resonance island in $(J,\hat\theta)$ space
ensures that the radial motion is localized.
So the remaining analysis should
concern the angular $(L,\phi)$ motion which reflects the slow evolution
of the Kepler ellipse. The structure of the $(L,\phi)$ space
influences values of quasi-energies as well as the structure of corresponding
semiclassical eigenstates. Trajectories in the $(L,\phi)$ space are determined
by constant values of $\Gamma$
also responsible for the size of the resonance island
in $(J,\hat\theta)$ space (recall that the island's size is determined 
 approximately  by $\sqrt{F\Gamma}$).

Before presenting the results let us define scaled variables, typically used
as a convenient parameterization of the microwave ionization problem (since
the dynamics scales classically \cite{perc}). The common choice is to link
the scaling to the initial energy of the electron \cite{KoLe95}.
 For an unperturbed 2D hydrogen atom the eigenenergies are given by
  $E_n=-1/2(n+1/2)^2$ with $n$ being non-negative
integer. This allows us to define scaled variables via the principal quantum
number of the initial state $n_0$ via
\bea
\omega_0&=&\omega(n_0+1/2)^3\\
F_0&=& F(n_0+1/2)^4\\
L_0&=&L/(n_0+1/2).
\eea
In particular note that the scaled angular momentum $L_0$ may take values
from $[-1,1]$ interval with extremal values corresponding to circular
orbits on which the electron moves in two opposite directions.

\section{Results and discussion}

To compare the semiclassical predictions to the exact quantum calculations,
we consider the $n_0=21$ manifold of our 2D model atom. 
This value is a compromise
between the requirement to be in the semiclassical, large $n_0$ regime
and computer memory limitations (size of the Floquet matrix to be diagonalized).
For resonant driving we take the microwave frequency to be
 $\omega=\omega_K=1/(n_0+1/2)^3$, ($\omega_0=1$).
  
In Fig.~\ref{qep1} we present values of $\Gamma$ as a function of
the scaled angular momentum, $L_0$
 and the $\phi$ angle for two different values of
the degree of the field ellipticity given by $\alpha$. Equipotential curves of
$\Gamma$ corresponding to semiclassical states originating from
$n_0=21$ hydrogenic manifold are also shown in the Figure. Due to the
specific form of the resonance Hamiltonian (\ref{EPr}) those lines are
{\it independent} of the microwave amplitude, $F$.

For $\alpha=0$ the $(L,\phi)$ space is symmetric with respect to the
$L_0=0$ axis since, in the LP case, dynamics is not
affected by the direction of the rotation of an electron. The left column
 in Fig.~\ref{qep1} presents the results for $\alpha=0.1$, i.e.,
the case very close to the LP problem. Note the presence of
four stable fixed points: two of them,
$L_0\approx 0$ and $\phi=\pi/2,\ 3\pi/2$,
correspond to almost straight line orbits oriented perpendicularly to the
main axis of the polarization ellipse,
 the other two fixed points correspond to
circular orbits, $|L_0|\approx 1$
(for such orbits $\phi$ is a dummy variable), with an electron
rotating in the same or opposite direction to the direction of
the rotation of the field
vector. In the vicinity of these fixed points the electronic motion
is ``shape'' localized (since the motion remains in the vicinity of a given
$(L_0,\phi)$ point the eccentricity of the orbit, $e=\sqrt{1-L_0^2}$, as well
as its orientation with respect to the polarization ellipse, given by $\phi$ 
is approximately preserved). Whether the motion is well localized in
its radial motion in $(J,\hat\theta)$ variables, i.e. along the ellipse 
depends on the size of the resonance island, given by $\sqrt{F\Gamma}$
as mentioned above. If the island in $(J,\hat\theta)$ space is sufficiently
large (comparable in size to $\hbar$ or larger)
 the quantum state will be localized in the island. On the contrary,
 too small resonance island cannot lead to localization in $(J,\hat\theta)$
 space, possible quantum states will spread over the whole ellipse
 (then also the semiclassical approach used becomes obviously non
 adequate).
 Thus to observe interesting, nonspreading wavepackets the resonance island
 must be large enough.
 
 In this respect, two fixed points corresponding for small $\alpha$ 
 to almost circular orbits, $|L_0|\approx 1$ lay at the maxima of
 $\Gamma$ (compare the left top panel in the figure) and for sufficiently large
 $F$ may enable
  a strong radial localization. The other two stable fixed points, corresponding
  to elongated orbits lay at local minima of $\Gamma$. 
 There exist also two unstable fixed points around $L_0\approx 0$ and
$\phi=0,\ \pi$, they form the origin of the separatrix dividing
the space.

When $\alpha$ increases the fixed points situated initially around
$L_0\approx 0$ and $\phi=\pi/2,\ 3\pi/2$ move in the direction of greater
negative values of $L_0$ (see the right column in Fig.~\ref{qep1}),
thus, the eccentricity of the corresponding orbits decreases.
The circular orbits do not change the shape but the resonance island
in $(J,\hat\theta)$ space
associated with  $L_0\approx 1$ becomes larger while that
associated with  $L_0\approx -1$ becomes smaller as reflected
by the greater and smaller values of $\Gamma$, respectively.
Note also that the islands in the $(L,\phi)$
space containing
librational states  shrink with the increase of $\alpha$, hence,
they can support
less and less semiclassical states -- while $\alpha$ increases
librational states vault over the separatrices and become rotational.
In the limit of $\alpha=1$ the islands disappear and there exist only
rotational states.

Consider the level dynamics corresponding to the change of $\alpha$.
Fig.~\ref{levd} shows the level dynamics for the group of states
originating from the $n_0=21$ hydrogenic manifold. We take the scaled
field amplitude
$F_0=0.03$, which for the CP case corresponds already to a significant
ionization yield \cite{ZGD96} and may be, therefore, considered as a 
quite large value.
 For the LP case,
i.e. $\alpha=0$, all states are doubly degenerate. Those of them which
correspond to rotational states in the $(L,\phi)$ space are
degenerate because the change of the rotation of an electron does
not affect dynamics in the LP problem [the $(L,\phi)$ space is symmetric
with respect to the $L_0=0$ axis]. The remaining librational states
are degenerate as the two islands, around $L_0=0$, $\phi=\pi/2$ and
$L_0=0$, $\phi=3\pi/2$, are identical and
support identical states. Of course
the degeneracies may be removed due to tunneling
processes, for instance, the wavepackets localized on circular orbits
(the highest levels) rotate in the opposite directions
and belong to distinct semiclassical states but quantum mechanical
states are the symmetric or antisymmetric linear combination of them.
Thus two wavepackets
propagating on a circular orbit in the opposite directions correspond to
a single eigenstate
(these wavepackets were already discussed in \cite{ke95} for the limiting
LP case only).
Similarly tunneling effects affect the librational states, i.e.
quantum states are the linear combinations of
solutions in each island.

An increase of $\alpha$ removes the semiclassical
degeneracy of rotational states as
one can see in Fig.~\ref{levd} -- observe splittings of upper levels in the
manifold. It corresponds to the broken symmetry with respect to
change of a direction of rotation of an electron. Quasi-energies of
states corresponding to the motion in the opposite sense to the rotation of
the field vectors, i.e. corresponding to negative values of $L_0$, move
down while those with positive $L_0$ go up. It is a consequence of
the behavior of $\Gamma$, i.e. the greater
$\Gamma$ the greater the quasi-energy value, see Figs.~\ref{qep1}.
Observe also that the degeneracy of all librational states (lower levels
in Fig.~\ref{levd}) is not immediately removed after a slight
change of the field ellipticity from $\alpha=0$ but is removed successively
during an increase of $\alpha$. This reflects the shrinking of the
corresponding
islands (the situation mentioned above) which causes the librational states
to vault over the separatrix and become rotational.
The levels with the smallest energy difference
correspond to the librational and
rotational orbits closest to the separatrix.
The narrowing of the level spacing in their vicinity
is just a consequence of the slowing down of the classical motion \cite{dd2}.
In the limit of $\alpha=1$ there is no degeneracy in the manifold.

Note that the energy splitting of the manifold
is the largest
in CP case while the smallest for LP.
It is associated with the corresponding strength
of the perturbation and simply expresses the dependence of $\Gamma$ on
$\alpha$. 

 Exact quantum results coming from the
numerical diagonalization of the Floquet Hamiltonian are presented also
in Fig.~\ref{levd}. Among the  multitude
of Floquet states appearing in the same energy range only those with the
largest overlap on the initial manifold are plotted for clarity.
One can see that
the agreement with the semiclassical predictions is very good except
in the region of broad avoided
crossings (with other levels - partners in the crossing eliminated by the
overlap selection) which appear in the upper part of the Figure.
 The semiclassical
method does not take into account the interaction of the considered
$n_0=21$ manifold with other manifolds (the method describes only
a single resonant
manifold), thus such avoided crossings have no chance to appear
in our semiclassical calculations.
For a more accurate comparison we plotted, in Fig.~\ref{energ}, quasi-energies
for two different values of $\alpha$ separately.
The agreement between the semiclassical
and quantum results is very good for high lying levels
(greater values of $\Gamma$) while for lower levels some differences
appear as expected due to small size of the resonance island.

In the limiting case $\alpha=1$, for CP case, another analysis of resonant
dynamics is possible, namely a harmonic approximation around a stable
equilibrium point in the rotating frame \cite{ibb} (corresponding to the
expansion about a stable fixed point $L_0=1$, $\phi$ arbitrary in our picture).
It is interesting to compare both approaches. The present semiclassical
quantization of the resonance hamiltonian is certainly a more general approach
valid for a general EP case and for the whole manifold. The harmonic
expansion is limited to the CP case  and valid for highest lying states
in the manifold only. On the other hand this expansion is quadratic in
deviations from the equilibrium point but nonlinear in the microwave field $F$
strength while the resonance Hamiltonian approach is a first order in $F$
expansion. Clearly in the deep semiclassical regime (large $n_0$) and for
sufficiently large $F_0$ the harmonic expansion approach yields a better
approximation for wavepacket states in the CP than the present resonant
Hamiltonian analysis (yielding, however, little information on the energies
of other states in the resonant manifold). On the other hand, quite
surprisingly, we have found, by a direct comparison of numerical values,
that for $F_0$ around 0.01 or 0.06 and $n_0$ up to 30-40 the resonant
Hamiltonian yields semiclassical values closer to exact quantum data than
the harmonic expansion. Thus for intermediate range of $n_0$ and $F_0$ values
the resonant Hamiltonian quantization is surprisingly good in accuracy,
yielding, at the same time, the predictions for the whole resonant manifold and
for an arbitrary polarization.

To complete the picture we show
in Fig.~\ref{levd1} the level dynamics of the same $n_0=21$ manifold
versus scaled field amplitude, $F_0$, for $\alpha=0.6$.
Quantum results are presented
together with the semiclassical ones. Again, for the
resonance island in orbital $(J,\hat\theta)$
motion sufficiently large to capture
quantum states, semiclassical results reproduce quantum ones
quite well even beyond a classical chaos border (of course
except avoided crossings).

Exact quantum Floquet matrix diagonalizations, performed using the Lanczos code
yields not only the eigenvalues but also the corresponding eigenvectors.
Their time evolution may be visualized \cite{abgre} to confirm 
directly that
indeed the Floquet states localized both in $(J,\hat\theta)$ and in
$(L,\phi)$ spaces correspond to the nonspreading localized wavepackets.
Consider again the level dynamics presented in Fig.~\ref{levd}. As
discussed above the higher lying state in the manifold corresponds to the
wavepacket motion of the electron on the circular orbit in the direction 
coinciding with the direction of rotation of the electric field vector.
And indeed such a motion is revealed by the plots of the corresponding
Floquet state (see Fig.~\ref{packets}, upper row). This Floquet state
changes little with change of the ellipticity of the microwaves, $\alpha$,
and in the limiting case of CP becomes a well known CP nonspreading wavepacket
\cite{ibb,dzb95}. In the opposite limit of $\alpha=0$ i.e. the LP microwave case
this state becomes almost degenerate with another one corresponding to
the different direction of the electron rotation. The two, almost
degenerate exact Floquet states are linear combinations of the two wavepackets
(at least in 2D) as discussed in \cite{ke95}.

It seems interesting to see what happens to the second member of the pair
as polarization is changed from being linear to elliptical
($\alpha$ increases). The two states separate fast in
$\alpha$ and {\it each} of them represents a distinct motion. As mentioned
before the state going down in energy corresponds to wavepacket moving
on the circular orbit in the direction opposite to the field. It undergoes 
a series of avoided crossings with other states of the manifold loosing
progressively its localized character. Still, for $\alpha$ not too large,
and far from avoided crossings its wavepacket character is clearly visible
(compare lower row in Fig.~\ref{packets}). This state looses its
wavepacket character when the librational islands (compare Fig.~\ref{qep1})
move sufficiently far down so as  rotational states with large negative
$L_0$ disappear since there is no ``space left'' for them due to the finite
value of $\hbar$.  
As may be seen by comparison of Fig.~\ref{qep1} and Fig.~\ref{levd} this
transition occurs around $\alpha=0.45$ for $n_0=21$, manifesting itself
in the quantum data (Fig.~\ref{levd}) 
by relatively broad avoided crossings encountered by
the wavepacket state. The transition point is, of course, $\hbar$ dependent, for
smaller $\hbar$ (larger $n_0$) the rotational, wavepacket-like, state with
large negative $L_0$ exists for larger $\alpha$. 
It is worth stressing that for $\alpha$ below this critical, $n_0$
(and thus driving frequency) dependent value there are two distinct Floquet
states corresponding to two wavepackets moving in opposite directions.
 
\section{Conclusions}

To conclude we have shown, by the comparison with the quantum numerical
results, that the proposed semiclassical method, based on the first order
resonant Hamiltonian  gives good quantitative
predictions. Numerical calculations produce nothing but
numbers and an advantage
of a semiclassical analysis is an understanding what physics is hidden
behind them. Apparently the behavior of H atom placed in EP microwave field,
in the range of parameters used above, is determined by the underlying
classical dynamics, especially, one may build wavepackets which follow
classical trajectories without spreading.
We have shown that the wavepacket discovered in CP microwaves
\cite{ibb,dzb95} corresponding
to the motion of the electron on a circular orbit in the same direction
as the rotation of the microwave field exists also in EP case. For 
a sufficiently small ellipticity of the microwave polarization, $\alpha$,
there exist also another wavepacket state, corresponding to electron 
rotating in the opposite direction to the microwave field.  
In the limiting LP case both wavepackets coalesce, the exact Floquet states
correspond semiclassically to linear combinations of  two wavepackets
propagating in the opposite direction on the circular orbit 
\cite{ke95}.

The analysis presented is restricted to the two-dimensional model atom
(similar assumption is implicit in \cite{ke95}), its validity for the
real three-dimensional atom is an open question. Certainly, in the limiting
LP case, due to rotational symmetry with respect to the field axis, 
the two wavepackets moving in the
opposite directions in 2D lead in 3D to a doughnut shaped localized
function oscillating between north and south pole of the sphere (assuming
a vertical polarization of LP microwaves) \cite{bsdz98}.
 For CP, on the other hand,
the wavepacket motion remains essentially two-dimensional \cite{zdb95}.
It will be most interesting to see how the third dimension affects the
dynamics also for a general case of EP microwave field since already
classical studies \cite{sz97,sz98} indicate that some qualitative
differences may appear. This subject is left for future studies. 

Still
the present 2D analysis shows that the nonspreading wavepacket states
localized on circular orbits are not restricted to circular polarization
of microwaves only. Thus the perfect circular polarization is not essential
for a possible experimental observation of nonspreading wavepackets. 

Interestingly, in the EP case,
there appears a possibility of the angular localization in minima of
the effective potential $\Gamma$ with the position of the minima being
$\alpha$ dependent (compare Fig.~\ref{qep1}).
 If the creation of localized wavepackets in such minima
were possible it would allow to control the shape of the trajectories
on which the wavepackets propagate by a change of microwave polarization,
and not by an additional static field as proposed in \cite{szd98}.
Study of the corresponding wavefunctions indicates, however, that
due to the small resonance island width the localization {\it along}
the ellipse [i.e. in ($J,\hat\theta$) space] is not very effective
at least in the range of $n_0$ and $F_0$ studied by us. 


\section{Acknowledgments}
We are grateful to Dominique Delande for numerous discussions
and the possibility to use his Lanczos diagonalization routines.
K.S. is a fellow of the Foundation for Polish Science.
Support of KBN
under project No.~2P302B-00915 (K.S. and J.Z.) is acknowledged.
Numerical calculations were performed at Academic Computer Center Cyfronet
in Krak\'ow with the help of grants KBN/S2000/UJ/067/1998 and
KBN/S2000/UJ/068/1998.
\section{Appendix}

Consider the Floquet Hamiltonian, $H_F$, as defined in (\ref{floq}) with
the hamiltonian given by (\ref{ham}).
For the sake of efficiency of the numerical diagonalization it is
convenient
to rewrite the Floquet Hamiltonian in the velocity gauge by applying
a standard unitary transformation \cite{cct,sth}
\be
H_F=-i\frac{\partial}{\partial t}+
\frac{p_x^2+p_y^2}{2}-\frac{1}{r}+\frac{F}{\omega}[p_x\sin\omega t-
   p_y\alpha\cos\omega t]
   +\frac{F^2}{4\omega^2}(\alpha^2+1).
\label{velo}
\ee
Next we pass to scaled semiparabolic coordinates (\ref{scaluv}) which
allows us to express the
Floquet eigenvalue equation
\be
H_F\mid u_n(t)\rangle=E_n\mid u_n(t)\rangle
\label{eigen}
\ee
as a generalized eigenvalue problem
\be
({\cal A}-E_n{\cal B})\mid u_n(t)\rangle=0
\label{eigen1}
\ee
where
\be
{\cal A}=(u^2+v^2)H_F \ \ \  \mbox{and} \ \ \ {\cal B}=u^2+v^2
\ee
with
\be
H_F=\frac{p_u^2+p_v^2}{2\Lambda^2(u^2+v^2)}-\frac{2}{\Lambda(u^2+v^2)}
-i\frac{\partial}{\partial t} +\frac{F}{\Lambda\omega (u^2+v^2)}[
(up_u-vp_v)\sin\omega t-\alpha (vp_u+up_v)\cos\omega t].
\ee
(with the additive pondoromotive term omitted).

Floquet states are periodic in time
(with the period $T=2\pi/\omega$) thus we may expand $\mid u_n(t)\rangle$
in the Fourier series
\be
\mid u_n(t)\rangle=\sum_{K=-\infty}^{+\infty}e^{-iK\omega t}\mid u_n^K\rangle
\ee
This allows to cast Eq.~(\ref{eigen1}) into the equivalent set of coupled
equations

\be
\left[\frac{p_u^2+p_v^2}{2\Lambda^2}-\frac{2}{\Lambda}-K\omega
(u^2+v^2)\right]\mid u_n^K\rangle 
+(W_s-W_c)\mid u_n^{K-1}\rangle
-(W_s+W_c)\mid u_n^{K+1}\rangle  =
E_n(u^2+v^2)\mid u_n^K\rangle
\label{seteq}
\ee
where
\bea
W_s&=&i\frac{F}{2\omega\Lambda}(up_u-vp_v) \cr
W_c&=&\alpha\frac{F}{2\omega\Lambda}(vp_u+up_v).
\eea
All terms in the above equation have the polynomial form in coordinates
and momenta. This suggests to use the harmonic oscillator basis
for an efficient evaluation of matrix elements \cite{sth}. The method becomes
then analogous to the treatment of the circular polarization case discussed
in detail elsewhere \cite{ZGD96}. The resulting generalized eigenvalue
equation is diagonalized using the Lanczos code which allows for
an extraction of eigenvalues in a selected energy range and the corresponding
eigenvectors. For completeness let us mention only that
the Floquet Hamiltonian is invariant under the generalized parity
transformation, i.e. the parity transformation combined with the
translation in time by $\pi/\omega$.
 Thus ${\cal A}$ and ${\cal B}$
matrices may be split into uncoupled matrices that are two times
smaller. This  makes
the numerical calculations more efficient.


\begin{figure}
\caption{Two dimensional hydrogen atom illuminated by resonant,
$\omega_0=1$,
elliptically polarized microwaves. The effective
scaled perturbation
$\Gamma_{0}=\Gamma/(n_0+1/2)^2$ is 
plotted as a function of
the scaled angular momentum, $L_0$, and the angle, $\phi$, between the
Runge-Lenz vector and the main axis of the polarization ellipse -- upper row.
Bottom row shows equipotential curves of the angular part $\Gamma$ of
the Hamiltonian ${\cal H}_r$, Eq.~(\protect{\ref{EPr}}), representing the slow
evolution of the Kepler ellipse. The curves correspond to semiclassical
states originating from the $n_0=21$ hydrogenic manifold.
Columns correspond to the different ellipticity of the microwaves
$\alpha=0.1$ (left) and $\alpha=0.6$ (right).
}
\label{qep1}
\end{figure}
\begin{figure}
\caption{Two dimensional hydrogen atom driven by resonant,
$\omega_0=1$,
elliptically polarized microwaves.
Level dynamics, versus $\alpha$ (i.e. the degree of the field
ellipticity), of the semiclassical quasi-energies (full lines) of the
states originating from the $n_0=21$ hydrogenic manifold for $F_0=0.03$ 
 compared with the exact quantum results (dotted lines).
}
\label{levd}
\end{figure}

\begin{figure}
\caption{
Comparison of the semiclassical quasi-energies (circles)
originating from the unperturbed $n_0=21$ manifold for $F_0=0.03$ and
$\omega_0=1$
with the exact quantum
values (crosses), at different values of the degree of the field ellipticity
$\alpha=0.1$ (a), $0.9$ (b). Integer index p counts consecutive states in the
perturbed manifold.
}
\label{energ}
\end{figure}

\begin{figure}
\caption{
Level dynamics of the exact quantum quasi-energies (dotted lines)
in the vicinity of the resonant manifold emerging from $n_0=21$,
compared with the semiclassical prediction (full lines), for $F_0=0\ldots 0.06$
and $\alpha=0.6$, $\omega_0=1$.
Note that the maximum field amplitude clearly exceeds typical ionization
thresholds measured in current experiments at the principal
resonance. The semiclassical prediction accurately
tracks the exact
solution across a large number of avoided crossings.
}
\label{levd1}
\end{figure}

\begin{figure}
\caption{Nonspreading wavepackets for 2D hydrogen atom illuminated by
the elliptically polarized microwaves with the amplitude $F_0=0.03$,
 frequency
$\omega_0=1$ for $n_0=21$ and ellipticity
$\alpha=0.4$. Top row -- the exact Floquet state corresponding to a
nonspreading wavepacket moving on a circular orbit in the direction
of the rotation of the microwave field at times $\omega t=0,\pi/4,\pi/2$ 
from left
to right. This wavepacket corresponds to the one known for the circular
polarization and may be obtained from the latter by a 
 change of the
microwave polarization. Bottom row -- another exact Floquet state
corresponding to a nonspreading wavepacket moving in the opposite direction
(shown at the same times $\omega t =0,\pi/4,\pi/2$ from left to right).
Note that while the former almost preserves its shape in temporal evolution,
the latter becomes significantly distorted. Still being an exact Floquet
time-periodic eigenstate it regains its shape (as depicted e.g. in the bottom 
left corner of the figure) every period of the microwave.
The size of each box is $\pm 800$ Bohr radii in both $x$ and $y$ directions. 
}
\label{packets}
\end{figure}

\end{document}